\documentclass[letter]{aa}  
\usepackage{graphicx}
\usepackage{xcolor}
\usepackage{ulem}
\usepackage{multicol}
\usepackage{bm}
\usepackage{txfonts}

\def\msun{\ifmmode M_{\odot} \else M$_{\odot}$\fi}
\def\msunyr{\ifmmode M_{\odot} {\rm yr}^{-1} \else M$_{\odot}$ yr$^{-1}$\fi}
\def\zsun{\ifmmode Z_{\odot} \else Z$_{\odot}$\fi}
\def\lsun{\ifmmode L_{\odot} \else L$_{\odot}$\fi}
\newcommand{\oh}{\ifmmode 12 + \log({\rm O/H}) \else$12 + \log({\rm O/H})$\fi}

\def\hii{H{\sc ii}}
\def\Oii{[O~{\sc ii}] $\lambda$3726,3729}
\newcommand{\Niiiuv}{N~{\sc iii}] $\lambda$1750}
\newcommand{\Nivuv}{N~{\sc iv}] $\lambda$1486}

\usepackage{soul}

\begin{document} 
   \title{N-enhancement in GN-z11: First evidence for supermassive stars nucleosynthesis in proto-globular clusters-like conditions at high redshift ?}

   \subtitle{}

\author{
   C. Charbonnel\inst{1,2}
   \and
   D. Schaerer\inst{1,2}
   \and
   N. Prantzos\inst{3}
   \and
   L. Ram\'irez-Galeano\inst{1}
   \and
   T. Fragos\inst{1}
   \and
    A. Kuruvanthodi\inst{1}
   \and
   R. Marques-Chaves\inst{1}
   \and M. Gieles\inst{4,5} }
\institute{
Department of Astronomy, University of Geneva, Chemin P\'egasi 51, 1290 Versoix, Switzerland \\
\email{corinne.charbonnel@unige.ch}
	 \and 
	 IRAP, UMR 5277 CNRS and Universit\'e de Toulouse, 14, Av. E.Belin, 31400 Toulouse, France 
  \and
  Institut d'Astrophysique de Paris, UMR 7095 CNRS, Sorbonne Université, 98bis, Bd Arago, 75014 Paris, France
 \and ICREA,  Pg. Llu\'{i}s Companys 23, E08010 Barcelona, Spain
   \and Institut de Ci\`{e}ncies del Cosmos (ICCUB), Universitat de Barcelona (IEEC-UB), Mart\'{i} i Franqu\`{e}s 1, E08028 Barcelona, Spain
}
   \date{Accepted by A$\&$A Letters}
  \abstract{Unusually high N/O abundance ratios were recently reported for a very compact, intensively star-forming object GN-z11 at $z=10.6$ from JWST/NIRSpec observations. We present an empirical comparison with the C, N, and O abundance ratios in Galactic globular clusters (GCs) over a large metallicity range. We show that hot hydrogen-burning nucleosynthesis within supermassive stars (SMS) formed through runaway collisions can consistently
  explain the observed abundances ratio in GN-z11 and in GCs.
This suggests that a proto-globular cluster hosting a SMS could be at the origin of the strong N-enrichment in GN-z11. Our model predicts the behavior of N/O, C/O, and Ne/O ratios as a function 
 of metallicity, which can be tested if  high-$z$ objects similar to GN-z11 are detected with JWST in the future. Further studies and statistics will help differentiate the proto-GC scenario from the Wolf-Rayet scenario that we quantify with a population synthesis model, and shed more light on this peculiar object.
  }


   \titlerunning{GN-z11: Evidence for supermassive stars nucleosynthesis in proto-GC conditions? }
   \authorrunning{C. Charbonnel et al.}
   \maketitle
%

\section{Introduction}
\label{section:introduction}
The detection of unusually bright NIII] and NIV] UV emission lines in the JWST/NIRSpec spectrum of GN-z11 \citep{2023arXiv230207256B}
revealed exceptional nitrogen enrichment in this high-$z$ galaxy. 
The N/O abundance ratio in the interstellar medium (ISM) of GN-z11  derived by \citet[][see also \citealt{Senchyna2023}]{2023arXiv230210142C}
is more than four times solar (log(N/O)$\ga -0.25$). This is more than one order of magnitude higher than what is usually found in low-redshift galaxies and \hii\ regions of similar low metallicity (12+log(O/H)$\la $8.0),
and even slightly higher than in galaxies with super-solar metallicities \citep[e.g.][]{Berg2019The-Chemical-Ev,2016MNRAS.458.3466V}. 
The C/O ratio of GN-z11 is compatible with normal values but poorly constrained \citep{2023arXiv230210142C}.
Reports of other peculiar objects showing indications for high N/O exist, at low and intermediate redshift \citep[see e.g.][]{James2009A-VLT-VIMOS-stu,Telles2014A-Gemini/GMOS-s,Villar-Martin2004Nebular-and-ste}.

The peculiar abundance ratios inferred in GN-z11 cannot be explained by classical stellar yields and standard galactic chemical evolution. However,  they are commonly found in  globular clusters (GC) which are known to form in the early Universe \citep[][for a review]{2019A&ARv..27....8G}.  
It is now firmly established that in each of these old, massive, and compact clusters, 
a large fraction of the stars (so-called second population, 2P) show different levels of N-enrichment
correlated to O- and C-depletion and to Na and Al enrichment (with the most extreme 2P stars exhibiting log(N/O)$\ga +0.4$), 
while the so-called 1st population (1P) stars have the same chemical composition
as field halo stars of similar metallicity (with typical  log(N/O)$\la -1$). From the nucleosynthesis point of view, there is considerable evidence that 2P GC stars formed out of a mixture between the original proto-GC gas and
the yields of hot  H-burning through the CNO cycle and the NeNa- and MgAl-chains that were 
 ejected by short-lived 1P stars (so-called polluters) 
 in their host proto-cluster \citep[][and references therein]{pci17}. 
No signature of high He-enrichment and of other nucleosynthesis processes such as triple$-\alpha$, s-process, and explosive nucleosynthesis
was found in the vast majority of the GCs, except in the most massive ones 
that might be the remnants of dwarf galaxy nuclei 
\citep{2021ApJ...923...22M}. 
The same chemical peculiarities were found in all the GCs of the Milky Way and of the Local Group where they have been looked for \citep{2017MNRAS.464.3636M,2022A&A...660A..88L}, as well as in extragalactic 
massive star clusters of intermediate-age down to $\sim$2~Gyr \citep{2018MNRAS.473.2688M,2019MNRAS.489L..80B}, pointing to similar early enrichment histories. 
However, they were never observed in less massive open clusters, implying a minimum cluster mass ($\gtrsim $ a few 10$^4$~M$_{\odot}$, M$_V$ $\lesssim$ -5; \citealt{2017A&A...607A..44B}) 
for their formation.  

Several scenarios invoking different types of polluters were proposed to explain  
GC abundance anomalies \citep[][for a review]{BL18}. The most recent observational clues point toward the  nucleosynthesis contribution of exceptional 1P polluters such as super-massive stars (SMS) as proposed by \citet{2014MNRAS.437L..21D}. There is no consensus how such a star forms, but an attractive  scenario in the GC context is that it forms via stellar collisions.
\citet{G18} showed that proto-star clusters hosting large number of protostars ($\gtrsim  10^6$)  accreting gas at a high rate ($\gtrsim 10^5$~M$_{\odot}$.Myr$^{-1}$) can undergo runaway collisions leading to the formation of SMS with masses between $\sim 10^3$ and $\sim10^5$ M$_{\odot}$ in 1 to 2 Myr, before two-body relaxation stops the contraction of the system.
Additionally, the SMS can be continuously rejuvenated by stellar colisions and thus process H at the required temperature to explain all the 
nucleosynthesis patterns while keeping its He content low, in agreement with GC photometric constraints \citep{2018MNRAS.481.5098M}. While spewing processed material via its strong wind, it can thus act as a conveyor belt and process much more material through hot H-burning than the maximum mass it reaches during the proto-GC contraction, 
solving the mass budget problem that plagues other models \citep{2017IAUS..316....1C}. 
Last but not least, this SMS formation channel could occur at any redshift in massive and compact enough 
star clusters, and therefore also in the compact core of GN-z11, which has a high stellar density, as will be discussed below.

In this Letter, we argue 
that the same scenario could potentially  explain the high log(N/O) abundance ratios in GN-z11 (Sect.~\ref{comparison:GCdataandSMS}).  We examine its compatibility with other properties of this high-$z$ object, and we discuss the alternative WR option  (Sect.~\ref{Discussion}). 
  We conclude and make predictions for future observations of high-$z$ galaxies (Sect.~\ref{Conclusions}).

\section{GN-z11: Abundance ratios as a result of CNO-processing in a SMS in proto-GC-like conditions ?}
\label{comparison:GCdataandSMS}

\subsection{Comparison with GC data}
\label{comparison:GCdata}

\begin{figure}
\includegraphics[width=9cm,angle=0]{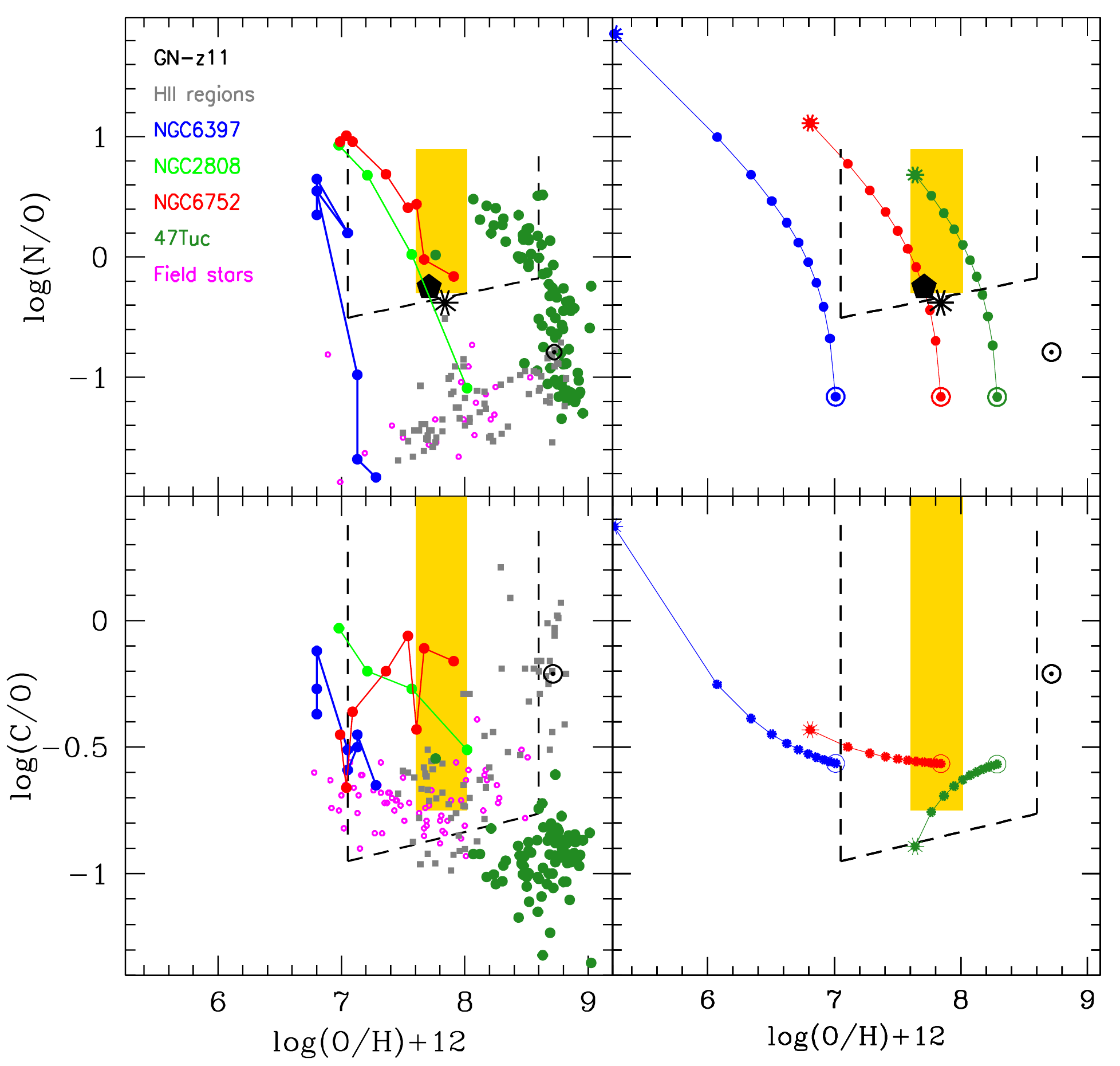}
\caption{
The yellow box and the black dashed line show the location of the abundance ratios of GN-z11 from  \citet{2023arXiv230210142C} fiducial model and from their more conservative assumptions respectively (with upper boundaries which are not well defined). The black pentagon and asterisk indicate the location of Senchyna et al. (2023) photoionization modeling results for GN-z11.
{\it Left:} 
Abundance ratios
in individual stars of 4 GCs (NGC~6397, \citealt{2005A&A...433..597C}; NGC~2808, \citealt{2023MNRAS.519.1695C}; 
NGC~6752,  \citealt{2005A&A...433..597C}, and 47~Tuc,  \citealt{2020MNRAS.492.1641M}), and in MW field stars within the  same metallicity range as the 4 GCs \citep{2004A&A...421..649I,2004A&A...414..931A,2009A&A...500.1143F}; for the GCs we connect the points by line segments to guide the eye. The grey squares are the abundance ratios for extragalactic HII regions and galaxies \citep[compilation from][]{Izotov_2023}.
{\it{Right:}} Theoretical 
abundance ratios as predicted from the dilution of the material of $10^4$~M$_{\odot}$ SMS models 
with the assumed initial proto-GC abundances for the same metallicities as NGC~6397, NGC~6752, and 47~Tuc (blue, red, and green respectively). 
Along each GC sequence, the fraction of SMS H-processed material in the mixture decreases from 1 (pure SMS, asterisk symbols) to 0 (pure proto-GC gas, open circles) with steps of 0.1 (full circles). 
}
\label{FigcompGCdata}
\end{figure}

To put our hypothesis in the GC observational context \citep[see also][]{Senchyna2023}, 
we chose four Galactic GCs  that cover a large range of metallicity and for which C, N, and O abundance measurements are simultaneously available,    
and we show in Fig.~\ref{FigcompGCdata} the values of 
 log(N/O) and log(C/O) as a function of log(O/H) in their individual stars and in MW field halo stars.  
 We compare to GN-z11 and other galaxies data.
 On one hand, each cluster contains 1P stars that present the same abundance ratios as Galactic 
 field stars and as 
low-redshift galaxies at the same metallicity. 
The typical log(N/O) of these objects ($\sim -1$ to $-2$ for 12+log(O/H)<8) is well explained by the production of primary N by fast-rotating massive stars of very low-metallicity \citep[][]{2006A&A...449L..27C,2017nuco.confa0401M}.  
On the other hand, the log(N/O) values in the 2P stars of the 4 GCs reach or even exceed the level observed in GN-z11.  
Similarly high log(N/O) ratios are ubiquitous in   
GCs, with the maximum values being found in the most metal-poor and$/$or the most massive GCs \citep{2020MNRAS.492.1641M}. 
Interestingly, the population of N-rich stars that has been found in the Milky Way inner halo \citep{2017MNRAS.465..501S} present other chemical peculiarities as GC 2P stars, and are likely  escapers from the GC population \citep{2023arXiv230303791G}.
The same agreement holds for log(C/O), although we note that no high values are found in 47~Tuc.

This comparison empirically supports  
the possibility that the bright N emission lines compared to faint O lines in GN-z11 may 
result from the  contamination of the gas of a massive and compact -- proto-GC-like -- star cluster  
by CNO-processed material from the same polluters as those required to explain GC abundance patterns
\citep[see also][]{Senchyna2023}. 

\subsection{Comparison with SMS nucleosynthesis}
\label{comparison:SMS}

\citet{G18} discussed how the maximum mass of a SMS formed via runaway collisions and the exact amount of H-processed material it ejects through winds depend on the uncertain stellar mass-radius relation and mass-loss rates adopted, as well as on the initial mass of the proto-GC where runaway collisions can occur. 
Using  realistic assumptions based for instance on the properties of very massive stars and the results of collisional $N$-body simulations, they show that for   
 the most massive proto-GC they consider and that hosts 10$^7$ proto-stars, the mass of the SMS can reach $\sim 5\%$ of the cluster stellar mass when the density of the cluster is maximum, and the mass it can process after 5~Myr is as high as $\sim 10^6$~M$_{\odot}$ (corresponding to $\sim 50 \%$ of the cluster stellar mass at that age).

To model  nucleosynthesis in fastly growing SMS,   
we computed evolution models for 3 metallicities ([Fe/H]=-0.72, -1.14, and -2~dex, assuming $[\alpha$/Fe]=+0.3~dex) with the stellar evolution code MESA \citep{Paxton2011,Paxton2013,Paxton2015,Paxton2018,2019ApJS..243...10P,Jermyn_2023}.
For a detailed description of our adopted stellar physics, we refer to \citet[][their Sect.3]{2023ApJS..264...45F}.  
We started from a low-mass seed (0.7~M$_{\odot}$) and applied high mass accretion rates to reach $10^4$~M$_{\odot}$ in $\sim$ 0.15~Myr after the runaway collision process started, as predicted in case of a proto-star cluster hosting $10^7$ stars \citep[for details see][]{G18}.
For this, we multiplied the prescription by \citet{2019A&A...624A.137H} for the accretion rate onto the star by a factor of 10.
In these conditions,  fresh H is permanently refilling the star and the central He-mass fraction stays close from its original value while the CNO-cycle is running at equilibrium.  The central temperature of the model thus  increases very slowly while the stellar mass growths (for the lowest metallicity considered, T$_c$ increases from 67 to 73~MK while the stellar mass increases from 3$\times 10^3$ to $10^4$~M$_{\odot}$). Hence, we find that accreting SMS satisfy the nucleosynthesis constraints over a large range of masses.
We did not include mass loss, but we assume that 
once mass accretion and mass loss roughly  compensate,  the conveyor belt for hot H-burning is fully efficient as predicted by \citet{G18}.
Hence, the chemical properties of the SMS ejecta
shall correspond to that of a "conveyor SMS" of the same mass. 
Depending on the initial number of proto-stars in the proto-GCs, the mass of the "conveyor SMS" and the amount of ejecta shall slightly vary, but their chemical properties will remain similar.

We show in  Fig.~\ref{FigcompGCdata} the predictions for the log(N/O) and log(C/O)  abundance ratios
as a function of log(O/H)+12 in the 
convective interior of the models 
when the stellar mass has reached 
10$^4$~M$_{\odot}$. 
The behavior of N/O versus O/H  differs from that of C/O when the SMS metallicity changes. While the N-enrichment results from the depletion of both C and O in the CNO cycle and occurs at all metallicity, C is relatively more depleted than O in higher metallicity SMS (for a given SMS mass, the central temperature is lower at higher 
metallicity). 
As a consequence, while the N/O ratio always increases in the SMS, with more extreme values due to stronger O depletion at lower metallicity, the C/O increases with respect to its initial value only in the most metal-poor case.  

For each metallicity, we  show in Fig.~\ref{FigcompGCdata} a sequence of points that correspond to the abundance ratios predicted  when the SMS material is mixed with the proto-GC gas in different proportions  
\citep[][]{2006A&A...458..135P,2014MNRAS.437L..21D}. 
The abundance ratios in GN-z11 require SMS ejecta being mixed with original proto-GC gas in proportions 
(region where the models overlap the boxes corresponding to GN-z11) that depend on the actual metallicity of GN-z11. This is in very good agreement with the dilution requirement to explain the abundance ratios in GCs, with  well-known variations from one GC to another (as seen among the 3 GCs in  Fig.~\ref{FigcompGCdata}), with the most extreme abundance ratios and the larger fraction of 2P versus 1P stars increasing with the GC mass and/or with decreasing metallicity
\citep{2017MNRAS.464.3636M,2019A&A...622A.191M}. This supports the super-linear correlation between the amount of mass that can be processed by a SMS and the mass of the cluster, as predicted by  \citet{G18}.  

As a final comment on nucleosynthesis, we recall that 2P stars in GCs also show high Na- and Al-enrichment. If our claim that the peculiar N, C, and O abundance ratios in   GN-z11 result from the same scenario as those in GCs, then its gas shall also be Na- and Al-enriched. Last but not least, it is mostly $^{22}$Ne that is involved in the NeNa-chain that leads to Na-enrichment, while the more abundant $^{20}$Ne is hardly changed in the SMS models. However, since O is strongly depleted, we predict that the Ne/O abundance ratio shall be higher in GN-z11 and other similar high-$z$ objects that shall be found in the future   compare to the case of normal galaxies of similar metallicities, with the highest Ne/O in the most metal-poor N-enriched objects. 

\section{Discussion}
\label{Discussion}

As just shown, the observed nebular abundances 
of GN-z11 are compatible with those of GCs and possibly due to CNO-processing in SMS formed via stellar collisions in massive and compact -- proto-GC-like -- star clusters. We now examine if other properties of this source are also compatible with this scenario and discuss possible alternatives.

\subsection{GN-z11: An extreme regime of massive star cluster polluted by SMS ?}

The size measurement and stellar mass estimate of \cite{2023arXiv230207234T}
translate to a compactness index 
$C_5 \ga (M_\star /10^5\msun)  (r_{\rm h}/pc)^{-1} \sim 200$, which is higher than that
of old Galactic GCs and young extragalactic massive star clusters by a factor of $\sim$ 30 to 15 respectively, as the comparison with \cite{2016A&A...587A..53K} shows. 
The same holds for the mass surface density ($\sim 4 \times 10^4$ \msun pc$^{-2}$) which is $\sim$ one order of magnitude higher than that of a typical old GC, but compatible with that derived for the high-z densest clouds identified in high-$z$ galaxies \citep{Claeyssens2023}.  
The stellar mass density (expressing the half-mass within the half-light radius) $\rho_\star \ga 600 $ \msun pc$^{-3}$ is also high 
 but not extreme compared to YMCs and GCs \citep{2018MNRAS.478.1520B}, and remains a lower limit, since the object is mostly unresolved. 
It is close to the initial stellar mass density considered by \cite{G18} for their GC formation scenario involving SMS.

Next we examine if sufficient amounts of enriched gas are available to explain the observed abundance ratios.
The total mass of enriched, ionized gas, which is directly observable, can easily be estimated assuming ionization equilibrium and a constant ISM density \citep[see, e.g.,][]{Dopita2003}:
\begin{equation}
        M_{\rm ionized} = \frac{m_p Q_H}{\alpha_B n_e} = 2.2 \times 10^7 \left(\frac{10^3}{n_e}\right) \left(\frac{Q_H}{8.8 \times 10^{54}}\right) \msun, 
\end{equation}
where $Q_H$ is the ionizing photon production rate which can be determined from H recombination lines, $n_e$ the electron density, $m_p$ the proton mass, and $\alpha_B$ the recombination rate coefficient. The main unknown here is the $n_e$, which cannot be determined using standard optical diagnostics (e.g.\ using \Oii\ doublet) from the available JWST observations of GN-z11, but which can be estimated from density-sensitive UV lines \citep{Senchyna2023}. 
The extreme compactness of GN-z11 clearly indicates higher-than-average densities. For example, the observed increase of $n_e$ with SFR surface density at $z\sim 1-3$  \citep{Reddy2023The-Impact-of-S} suggests a mininum of $n_e \gg 500$ cm$^{-3}$. Very high densities are possible and possibly expected \citep{2023arXiv230304827D}, but $n_e \gg 10^5$ cm$^{-3}$ is probably excluded
since the \Oii\ emission doublet (which has the lowest critical density of the lines observed) would then be significantly weaked \citep[][]{2023arXiv230207256B}. On the other hand, if the density is not uniform even higher electron densities could be reached in the core,
as suggested by \cite{Senchyna2023}. 

Adopting the ionization rate derived from observations \citep[$Q_H = 8.8 \times 10^{54}$ s$^{-1}$,][] {2023arXiv230207256B} and $n_e=10^5$ cm$^{-3}$, a plausible minimum mass of ionized gas is therefore $M_{\rm ionized} \sim 2 \times 10^5$ \msun. Assuming that the gas has a constant density and uniform abundances, this implies that the amount of enriched material should be of the same order of $\ga 10^5$ \msun, assuming a dilution factor of 0.5.
The scenario of \cite{G18} predict enriched ejecta (wind) masses between $\sim 10^4-10^6$ \msun\ which are ejected over $\sim 3$ Myr from the SMS. 
From the currently available data we therefore conclude that these amounts could be sufficient to significantly enrich the amount of ionized gas observed in GN-z11. 

Is the core of GN-z11 a single proto-GC-like object?
Both the inferred stellar mass ($M_\star{_{GN-z11}} \sim 10^9$ \msun) and the brightness of GN-z11 ($\sim 26$ AB magnitudes in the rest-UV) are fairly high, but not unseen of, compared to observed values for stellar clumps observed at high-reshift \citep[see][]{Claeyssens2023}. However, the exact nature of these clumps is not known, and we cannot assert whether these entities and the unresolved core of GN-z11 represent individual clusters (e.g.\ scaled-up versions of 30 Doradus) or more complex regions.
Clearly, for classical assumptions, GN-z11 is most likely too bright to host a single proto-GC-like cluster in the regime explored by \citet{Gieles18}. Assuming the UV emission of GN-z11 is dominated by a single young cluster of $\sim 5$ Myr age, the stellar mass would be $\sim 1.2 \times 10^8$ \msun, using the model of \cite{Boylan2017}.

Now, the IMF could favor more massive stars at high-$z$, for instance due to the rising CMB temperature, collisions in very dense media, or others \citep[e.g.][]{Steinhardt2022Implications-of}, in which case the inferred stellar mass would be lower.
The presence of an SMS could potentially further lower the stellar mass estimate. However, only for a maximum SMS mass of $M_{\rm SMS} \sim 10^6$ \msun\ explored by the scenario of \cite{G18}, the SMS would reach a magnitude of $m_{\rm AB} \sim 27.5$ at $z\sim 10$, and the cluster plus SMS then possibly $m_{\rm AB} \sim 27$, according to the predictions of \cite{Martins2020Spectral-proper}. It seems therefore difficult to explain the entire core of GN-z11 by a single SMS-hosting cluster, and we thus suggest that it contains other lower mass star clusters, possibly hosting multiple very massive stars and/or SMSs. 
Last, it could eventually fall in the regime of the most massive GCs like $\Omega$ Cen, M54, or NGC~1851, which present similar abundance patterns than  GCs, while being the remnants of dwarf galaxy nuclei. 

In any case, while these simple estimates leave some room for the simple toy-model of \cite{G18} scenario to explain the currently available observations of GN-z11, we cannot definitively conclude on the ``proto-GC" nature of this object. 

\subsection{... or by rotating massive star winds ? Alternative scenarios}

Different scenarios to explain the peculiar abundance pattern of GN-z11 have been discussed by \cite{2023arXiv230210142C}, albeit without quantitative estimates or models. 
These include the presence of a massive black hole, pollution from traditional evolved stars, the impact of primordial or exotic stellar evolution channels, tidal disruption events, or effects of stellar encounters in dense star clusters. None of these scenarios provide clear predictions, but for reasons discussed by \cite{2023arXiv230210142C}, they favor tidal disruption events or runaway stellar collisions in a dense stellar cluster.
\cite{Senchyna2023}, on the other hand, noting the similarity between the N/O abundances in GC stars and GN-z11, have proposed that this object could show possible signatures of GC precursors, similarly to our claim.

\begin{figure}
\begin{center}
\includegraphics[width=7cm,angle=0]{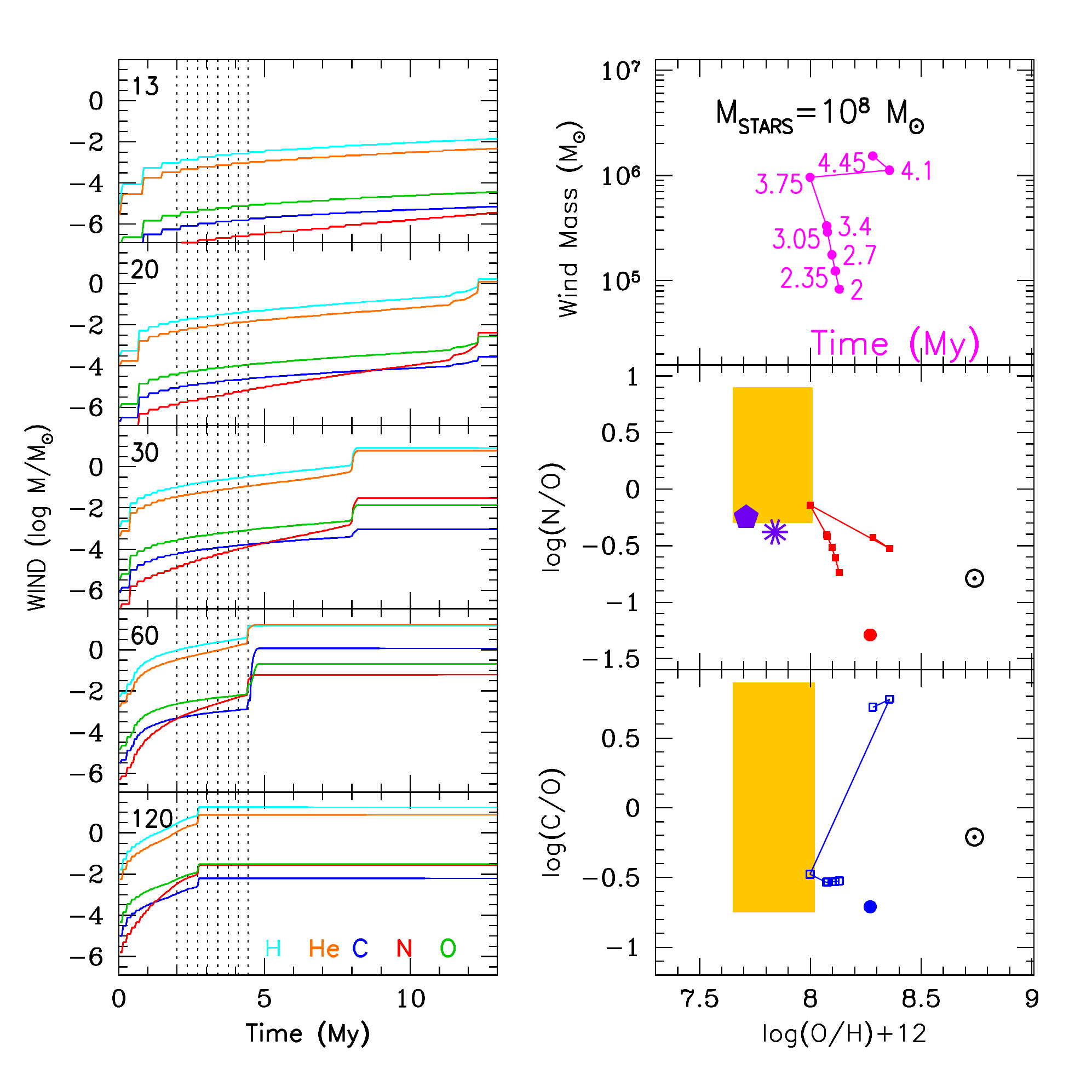}
\caption{{\it Left:} Cumulative mass in the winds of selected massive stars of Limongi and Chieffi (2018) with metallicity [Fe/H]=-1 and initial rotational velocity of 300 km/s (mass in M$_{\odot}$ on top left) as function of time. {\it Right Top}: Total mass ejected in the stellar winds of a starburst of total mass 10$^8$ M$_{\odot}$ with a Salpeter IMF at various times after the burst (as indicated on the left panels by dotted vertical lines) vs the O/H.  {\it Right Middle}: Corresponding evolution of the N/O when the ejecta are diluted in ambient gas of same mass; observational data as in Fig. 1.  {\it Right Bottom}: Same as in the middle but for C/O.
}
\label{Fig_Winds}
\end{center}
    \end{figure}

Local enrichment from massive star winds, presumably in the Wolf-Rayet (WR) phase, have also been discussed in both the above studies. To quantify this scenario, we have computed a population synthesis model using stellar wind yields of rotating massive stars of  metallicity [Fe/H]=-1 and initial rotational velocity of 300 km/s from  
\cite{Limongi2018}. Such yields have already been used in \cite{Prantzos2018} to successfully describe the chemical evolution of the solar neighborhood. The composition of the wind yields is shown on the left panels of Fig. \ref{Fig_Winds}. Assuming that stars more massive than 20-25 \msun   ~do not explode but collapse to black holes \citep[see][]{Prantzos2018}, the ejecta have exclusively wind composition for several Myr; they are dominated early on by H-burning products (N-rich, O- and C-poor, WN phase) and later by He-burning products (N-poor, C- and O- rich, WC phase).
Assuming a starburst of 10$^8$ \msun \ formed within 1 Myr with a Salpeter IMF, the mass ejected  after the first few Myr  is of the order of 10$^6$ \msun\ (Fig. \ref{Fig_Winds}, top right panel), that is  comparable to the gas mass estimates discussed above. 
For a short period, lasting a few 10$^5$ yr (as seen from the timeline in top right panel), the N/O and C/O values of the ejecta - assumed diluted to equal mass of original gas - reach the range of values reported for GN-z11; this period ends after $\sim$3.7 Myr (similar to the age reported in  \citealt{Senchyna2023}), when He-burning products lead to strong C and O enrichment of the stellar winds.

The above scenario is fairly simplistic (ignoring, among many other things, the role of binarity in the formation of WR stars)  and is presented for illustration purposes. In contrast to the SMS scenario, it is not related to GCs, since their defining feature (the Na-O anticorelation) cannot be produced by ``normal'' massive stars \citep{pci17}.
 Additionnally, it requires a rather special timing since the WN period ``favorable'' to enhance N/O is
 at best $\sim 1/10$ of the duration of the starburst, which is estimated to only a few Myr in GN-z11 \citep{2023arXiv230210142C,Senchyna2023}.

Local enrichment of N/O (and C/O in few cases) reaching the levels shown in Fig.\ \ref{Fig_Winds}, has been observed in some \hii\ regions hosting WR stars \citep[e.g.][]{Lopez-Sanchez2007The-Localized-C}. In general, however, and measured on larger scales, the N enrichment observed in WR galaxies is very modest, with average N/O enhancements of the order of $\sim 0.1$ dex only \citep{Brinchmann2008Galaxies-with-W}.
Furthermore, signatures of WR stars are not detected in GN-z11, or extremely weak at best, as pointed out by \cite{2023arXiv230207256B}, \cite{2023arXiv230210142C}, and \cite{Senchyna2023}. Finally, except for one peculiar galaxy showing WR features, Mrk 996, none shows so far the strong UV emission lines of \Niiiuv\ and \Nivuv\ observed in GN-z11 \citep[cf.][]{Senchyna2023}. 
The hypothesis of pollution from WR stars is therefore debatable.

In addition to the points discussed previously, other indirect arguments favoring the explanation of SMS caught in the act of polluting 
proto-GC-like massive star clusters might also be mentioned.
For example, it is known to require extremely compact and dense environments, as those found in GN-z11;   
the oldest GCs in the Milky Way have absolute ages of $\sim$ 13.6 $\pm$ 1.6 Gyr  \citep[][]{2017ApJ...838..162O},  making it possible to have GC-like forming conditions at very high $z$, even though the peak of GC formation is predicted to be at $z$ around 4 in  current hierarchical assembly models \citep[e.g.][]{2019MNRAS.482.4528E}. The interpretation of GN-z11 as GC progenitor may imply that these models and the role of GCs in reionisation need to be re-considered. 
On the other hand, WR
stars form in all conditions and at all times (of course with variable amounts depending on metallicity and other factors). Therefore abundance patterns resembling those of GCs stars should be found more frequently in ionized gas at high-$z$ and related to very compact sources. For a single object such ``statistical'' arguments are however, not properly applicable. Future observations and other studies should allow to further refine and test these and alternative scenarios.

\section{Conclusions}
\label{Conclusions}

In this Letter we present an empirical comparison of the C, N, and O abundance ratios in the compact $z=10.6$ galaxy GN-z11 derived recently from JWST observations with data in GC and field star observations, and a quantitative estimate of SMS nucleosynthesis that can form through runaway collisions in proto-GCs, as proposed in the scenario of \cite{G18}. We show that the SMS polluter models can simultaneously explain the observations of GN-z11 and of the GCs, with similar dilution values between the H-processed material in the SMS and the original proto-GC gas. Our model predicts that similarly high or even higher N/O and C/O ratios could be observed in proto-GCs in high-$z$ galaxies of lower metallicity than GN-z11, 
but that lower C/O ratios shall be expected 
for higher metallicities. On the other hand, the Ne/O ratios in similar objects shall be higher than in normal galaxies of similar metallicities, with an increase inversely proportional to the metallicity of the proto-GC host galaxy.
Finally, N-enriched and O-depleted gas in GN-z11 should also be enriched in Na and Al and in similar high-$z$ objects, if the observations reflect the conditions during GC formation.
Alternatively, the ``rotating massive star winds'' scenario could also explain the peculiar abundance patterns observed in GN-z11, but within a rather short time window, at least with current stellar models.  In view of the scarce current observational sample, this option cannot be totally excluded at present. However, in this case GN-z11 cannot be host of a proto-GC, as massive stars will later pollute their environment in He-burning products, at odds with observed abundances in GCs today. 

If the SMS scenario can be firmed up by future studies, this would provide an important step for our understanding of GCs and for the formation of SMS in general, with numerous important implications.
In any case, the peculiar properties of GN-z11 just revealed by JWST call for further studies to understand the physical processes ongoing in such extreme objects in the early Universe, and their possible connection with the formation of globulars, SMS, potentially also supermassive black holes and others \cite[see discussions in][]{2023arXiv230210142C,Senchyna2023}.
   
\begin{acknowledgements}
We thank the anonymous referee for constructive comments and Y.Izotov for providing data. CC acknowledges support from the Swiss National Science Foundation (SNF; Project 200020-192039).
LRG and TF acknowledge support from a 
SNF Professorship grant (Project No
PP00P2\_211006). MG acknowledges support from the Ministry of Science and Innovation (EUR2020-112157, PID2021-125485NB-C22) and from Grant CEX2019-000918-M funded by MCIN/AEI/10.13039/501100011033 and from AGAUR (SGR-2021-01069).
\end{acknowledgements}

 \bibliographystyle{aa} 
 \bibliography{Charbonnel_GN-z11} 
%

---------
 
\end{document}